\documentclass[pre]{revtex4}
\usepackage{amssymb}
\usepackage{amsmath}
\usepackage{graphicx}

\begin{document}

\title{Matter-wave soliton interferometer based on a nonlinear splitter}
\author{Hidetsugu Sakaguchi}
\affiliation{Department of Applied Science for Electronics and Materials,
Interdisciplinary Graduate School of Engineering Sciences, Kyushu
University, Kasuga, Fukuoka 816-8580, Japan}
\author{Boris A. Malomed}
\affiliation{Department of Physical Electronics, School of Electrical Engineering,
Faculty of Engineering, Tel Aviv University, Tel Aviv 69978, Israel}

\begin{abstract}
We elaborate a model of the interferometer which, unlike previously studied
ones, uses a local ($\delta $-functional) \emph{nonlinear} repulsive
potential, embedded into a harmonic-oscillator trapping potential, as the
splitter for the incident soliton. An estimate demonstrates that this
setting may be implemented by means of the localized Feshbach resonance
controlled by a focused laser beam. The same system may be realized as a
nonlinear waveguide in optics. Subsequent analysis produces an exact
solution for scattering of a plane wave in the linear medium on the $\delta $%
-functional nonlinear repulsive potential, and an approximate solution for
splitting of the incident soliton when the ambient medium is nonlinear. The
most essential result, obtained by means of systematic simulations, is that
the use of the nonlinear splitter provides the sensitivity of the
soliton-based interferometer to the target, inserted into one of its arms,
which is much higher than the sensitivity provided by the usual linear
splitter.
\end{abstract}

\maketitle

\section{Introduction and the model}

Matter-wave solitons are self-trapped modes in Bose gases with attractive
interactions between atoms, which have been created in Bose-Einstein
condensates (BECs) of $^{7}$Li \cite{Li} and $^{85}$Rb \cite{Rb,reflection}.
Available experimental methods make it possible to efficiently steer the
motion of solitons in matter-wave conduits, and study various dynamical
phenomena, such as reflection of solitons from potential barriers \cite%
{reflection} and collisions between solitons \cite{collision}.

\begin{figure}[tbp]
\begin{center}
\includegraphics[height=4.cm]{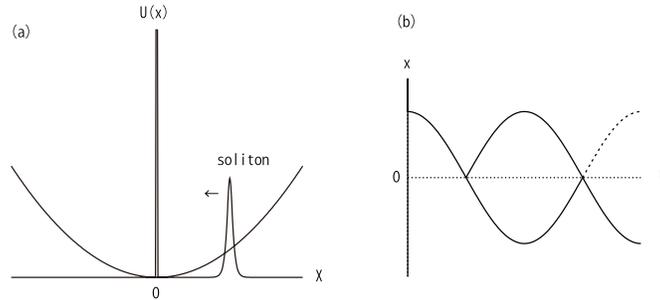}
\end{center}
\caption{(a) The scheme of the soliton-based interferometer, with the
initial soliton trapped in the HO potential and the splitting barrier
installed at the center. The arrow indicates the direction of motion of the
soliton. (b) A sketch of the operation of the interferometer, shown by means
of the spatiotemporal evolution of the density: splitting of the incident
soliton, followed by recombination of the secondary pulses (incomplete, in
the general case), after the collision between them.}
\label{f1}
\end{figure}

In addition to the obvious significance to fundamental studies, the solitons
may find an important application to the design of matter-wave
interferometers. Soliton-based interferometric schemes have been elaborated
in many theoretical works \cite{interf-theory}-\cite{Helm}, and recently
implemented in the experiment \cite{interf-exp}. The main element of the
interferometer is a narrow potential barrier, which provides for splitting
of an incident soliton into two matter-wave pulses, that move apart in the
harmonic-oscillator (HO) trapping potential into which the splitter is
embedded, and return back, to collide and recombine on the splitter, as
shown schematically in Fig. \ref{f1}.

The interferometric effect is produced by placing a target in one arm of the
interferometer (on the left- or right-hand side of the splitter), which
affects the outcome of the secondary collision by shifting the phase of the
pulse passing the target. The elaboration of this scheme makes it necessary
to study in detail collisions of the pulses with the potential barrier,
including such aspects as the deviation from one-dimensionality \cite{Cuevas}%
, finite width of the barrier \cite{wide-rectangular}, quantum effects
beyond the limits of the mean-field theory \cite{quantum,Helm,quantum2},
collisions of two-component solitons \cite{2comp-collision}, etc. For the
overall operation of the interferometer, a crucial factor is the phase
stability of the colliding pulses \cite{phase-control,collision}, as the
relative phase of the pulses determines the outcome of the collision. In
this respect, the use of the solitons offers a potential advantage, as its
collective phase degree of freedom is canonically conjugate to its norm (the
same is true for quantum states \cite{squeeze}), hence for heavy solitons
random phase fluctuations may be efficiently suppressed by the large value
of the norm.

The main objective of the present work is to elaborate the soliton-based
interferometric scheme which uses a\textit{\ \emph{nonlinear} }potential
barrier (splitter), instead of the linear repulsive defect studied in the
previous works. Assuming that the tight transverse confinement is provided
by an isotropic HO potential with frequency $\Omega _{\perp }$, the
corresponding mean-field wave function is looked for, as usual, in the
factorized form \cite{BEC}:%
\begin{equation}
\Psi \left( \rho ,X,T\right) =\frac{1}{\sqrt{\pi }a_{\perp }}\exp \left(
-i\hbar \Omega _{\perp }T-\frac{\rho ^{2}}{2a_{\perp }^{2}}\right) \Phi
\left( X,T\right) ,~a_{\perp }^{2}=\frac{\hbar }{m\Omega _{\perp }},
\label{Psi}
\end{equation}%
where $\rho $ and $X$ are, respectively, the transverse-radial and
longitudinal coordinates, $T$ is time, and $m$ the atomic mass. The ensuing
scaled form of the one-dimensional (1D) Gross-Pitaevskii equation (GPE),
which includes the barrier combining linear and nonlinear localized
repulsive potentials, with respective strengths $\varepsilon _{0}>0$ and $%
\varepsilon _{2}>0$, as well as the longitudinal HO potential with strength $%
\Omega ^{2}$, is
\begin{equation}
i\frac{\partial \phi }{\partial t}=\left[ -\frac{1}{2}\frac{\partial ^{2}}{%
\partial x^{2}}-g|\phi |^{2}+\frac{1}{2}\Omega ^{2}x^{2}+\varepsilon
_{0}\delta (x)+\varepsilon _{2}\delta (x)|\phi |^{2}\right] \phi ,
\label{GPE}
\end{equation}%
where the strength of the background self-attraction is set to be $g=1$,
unless $g=0$ in the linearized model, and $\delta (x)$ is the Dirac's delta
function. In numerical simulations reported below, it is replaced by a
rectangular potential tower of width $\Delta x$ and height $1/\Delta x$,
centered at $x=0$. The relation between the variables measured in physical
units and their scaled counterparts is%
\begin{equation}
X=X_{0}x,~T=\left( mX_{0}^{2}/\hbar \right) t,~\Phi =a_{\perp }\left( X_{0}%
\sqrt{\left\vert 2\left( a_{s}\right) _{\mathrm{backgr}}\right\vert }\right)
^{-1}\phi ,  \label{scaled}
\end{equation}%
where $X_{0}$ is a characteristic scale of the longitudinal coordinate, and $%
\left( a_{s}\right) _{\mathrm{backgr}}<0$ is the scattering lengths of
atomic collisions far from the barrier. Further, the frequency of the HO
potential, measured in physical units, is $\omega =\left( \hbar
/mX_{0}^{2}\right) \Omega $, the linear potential barrier is $\left( \hbar
^{2}/mX_{0}\right) \varepsilon _{0}\delta (X)$, and the nonlinear barrier is
defined, in terms of the full underlying GPE, as $\left( 4\pi \hbar
^{2}/m\right) X_{0}^{2}\varepsilon _{2}\delta (X)\left\vert \Psi \right\vert
^{2}$.

The Hamiltonian (energy) corresponding to Eq. (\ref{GPE}) is%
\begin{gather}
E=\frac{1}{2}\int_{-\infty }^{+\infty }\left( \left\vert \frac{\partial \phi
}{\partial x}\right\vert ^{2}-g\left\vert \phi \right\vert ^{4}+\Omega
^{2}x^{2}\left\vert \phi \right\vert ^{2}\right) dx  \notag \\
+\varepsilon _{0}\left\vert \phi (x=0)\right\vert ^{2}+\frac{1}{2}%
\varepsilon _{2}\delta (x)\left\vert \phi (x=0)\right\vert ^{4}  \label{H}
\end{gather}

The number of atoms in the condensate, given by the norm of wave function (%
\ref{Psi}) in physical units, is
\begin{equation}
N_{\mathrm{at}}=2\pi \int_{0}^{\infty }\rho d\rho \int_{-\infty }^{+\infty
}dX~\left\vert \Psi \left( \rho ,X\right) \right\vert ^{2}\equiv \pi
a_{\perp }^{2}\int_{-\infty }^{+\infty }\left\vert \Phi (x)\right\vert
^{2}dX.  \label{Nat}
\end{equation}
It is proportional to the norm of the scaled 1D wave function,
\begin{equation}
N=\int_{-\infty }^{+\infty }\left\vert \phi (x)\right\vert ^{2}dx.
\label{Nscaled}
\end{equation}%
Indeed, it follows from Eq. (\ref{scaled}) that
\begin{equation}
N_{\mathrm{at}}=a_{\perp }^{2}\left( 2X_{0}\left\vert \left( a_{s}\right) _{%
\mathrm{backgr}}\right\vert \right) ^{-1}N.  \label{N}
\end{equation}%
Estimates for typical values of physical parameters, including $N_{\mathrm{at%
}}$, which are relevant in the present context, are given below.

It is worthy to note that, for a sufficiently dense atomic gas, the
factorization procedure leads to a deviation of the effective 1D
nonlinearity from the simple cubic term \cite{1D3D}. The consideration of
the nonlinear barrier combined with the more sophisticated background
nonlinearity is an interesting issue too, which is left beyond the framework
of the present work.

We will chiefly consider the model with the fully nonlinear potential
barrier, setting $\varepsilon _{0}=0$ in Eq. (\ref{GPE}). Recently, soliton
dynamics in systems with spatially modulated nonlinearity has drawn much
attention, as it opens new possibilities for controlling outcomes of the
evolution of solitons by means of their amplitude, which is proportional to $%
N_{\mathrm{at}}$, see review \cite{RMP} and references therein. A still more
recent use of nonlinear potential barriers, of the same type as introduced
in Eq. (\ref{GPE}), was proposed in Ref. \cite{students} in a model of a
pumped laser cavity, where nonlinear barriers confined intra-cavity
solitons, but allowed the release of small-amplitude radiation, thus
stabilizing the trapped soliton modes \cite{students}. In optics, strong
local change of the nonlinearity may be induced by doping the respective
small-area region by atoms providing resonant two-photon interaction with
the electromagnetic wave \cite{Kip}. In this connection, it is worthy to
note that GPE (\ref{GPE}) may also be realized, in terms of optics, as the
nonlinear Schr\"{o}dinger equation for the spatial-domain propagation in a
planar waveguide, with transverse coordinate $x$, and $t$ replaced by the
appropriately scaled propagation distance, $z$. In that case, the trapping
potential defines the guiding channel, which is split into two by the $%
\delta $-functional terms $\sim \varepsilon _{0,2}$ \cite{NatPhot}.

For the atomic BEC, the localized barrier can be created by means of the
optically-controlled Feshbach resonance (FR) \cite%
{Feshbach-optical,Tom,Nicholson}, that reverses the intrinsic BEC\
nonlinearity from uniformly attractive to strongly repulsive in a narrow
region of width $\Delta x$, onto which the control laser beam is focused. It
is relevant to estimate physical parameters which will make the creation of
such a nonlinear barrier possible. Far from the barrier, the free soliton
with amplitude $A$, moving at velocity $v$, is given by the commonly known
solution of Eq. (\ref{GPE}) with $g=1$ and $\Omega =0$:%
\begin{equation}
\phi =A~\mathrm{sech}\left( x-\xi (t)\right) \exp \left( \frac{i}{2}\left(
A^{2}-v^{2}\right) t\right) ,~v=\frac{d\xi }{dt}.  \label{sol}
\end{equation}%
The scaled norm and energy of the free soliton are given by Eqs. (\ref%
{Nscaled}) and (\ref{H}) (dropping terms $\sim \Omega $ and $\varepsilon
_{0,2}$ in the latter equation), with $\phi (x)$ substituted by expression (%
\ref{Esol}):%
\begin{equation}
N_{\mathrm{sol}}=2A,~E_{\mathrm{sol}}=-(1/3)A^{3}+Av^{2},  \label{Esol}
\end{equation}%
its effective mass being $2A$, in the scaled notation. Thus, the amplitude
of the soliton is determined by the number of atoms bound in it ($N_{\mathrm{%
at}}$), as per Eqs. (\ref{Esol}) and (\ref{N}).

As shown below, in the framework of Eq. (\ref{GPE}) the soliton, hitting the
nonlinear barrier, may split into secondary pulses under condition $%
A\varepsilon _{2}\gtrsim 1$, see Eq. (\ref{min}). Undoing the above
rescalings, it is easy to convert the latter condition into one written for
physical parameters:%
\begin{equation}
\frac{\left( a_{s}\right) _{\mathrm{barrier}}}{\left\vert \left(
a_{s}\right) _{\mathrm{backgr}}\right\vert }\gtrsim \frac{a_{\perp }^{2}}{N_{%
\mathrm{at}}\left\vert \left( a_{s}\right) _{\mathrm{backgr}}\right\vert
\Delta X},  \label{ratio}
\end{equation}%
where $\left( a_{s}\right) _{\mathrm{barrier}}>0$ is the scattering lengths
of the atomic collisions switched by means of the FR inside of the barrier,
and $\Delta X\equiv X_{0}\Delta x$ is the width of the barrier in physical
units, whose minimum size, admitted by the diffraction limit for the control
optical beam, is $\Delta X\sim 1$ $\mathrm{\mu }$m. For instance, in the
case of the gas of $^{88}$Sr atoms, where the optically-controlled FR may be
used efficiently \cite{Tom}, the background scattering length is $\left(
a_{s}\right) _{\mathrm{background}}=-1.6\times \left( \mathrm{Bohr~radius}%
\right) \sim -0.1$ nm. Then, taking an appropriate value of the
transverse-confinement radius, $a_{\perp }\sim 1$ $\mathrm{\mu }$m (it
corresponds to $\Omega _{\perp }\sim 2\pi \times 100$ Hz), and $\Delta X\sim
1$ $\mathrm{\mu }$m, Eq. (\ref{ratio}) amounts to%
\begin{equation}
\frac{\left( a_{s}\right) _{\mathrm{barrier}}}{\left\vert \left(
a_{s}\right) _{\mathrm{backgr}}\right\vert }\gtrsim \frac{10^{4}}{N_{\mathrm{%
at}}}.  \label{1/N}
\end{equation}%
Available experimental techniques make it definitely possible to make the
ratio on the left-hand side of Eq. (\ref{1/N}) $\sim 10$, hence the
nonlinear barrier should work for solitons with $N_{\mathrm{at}}\gtrsim
10^{3}$, this number of atoms bound in the soliton being a realistic one. It
is also relevant to write the corresponding estimate for the axial size of
the soliton, corresponding to the above-mentioned values $a_{\perp }\sim 1$ $%
\mathrm{\mu }$m and $\left( a_{s}\right) _{\mathrm{backgr}}\sim 0.1$ nm, in
physical units:
\begin{equation}
l\sim \frac{a_{\perp }^{2}}{N_{\mathrm{at}}\left\vert \left( a_{s}\right) _{%
\mathrm{backgr}}\right\vert }\sim \frac{10^{4}}{N_{\mathrm{at}}}~\mathrm{\mu
m}.  \label{l}
\end{equation}%
In particular, for $N_{\mathrm{at}}\sim 10^{4}$, which is also a realistic
number of atoms in the soliton,\ Eq. (\ref{l}) yields $l\sim 1~\mathrm{\mu m}
$. In the combination with $a_{\perp }\sim 1$ $\mathrm{\mu m}$, this implies
that the soliton will effectively look as a nearly isotropic 3D object, in
agreement with its actual shape observed in the experiments \cite{Li}.

The detailed theoretical analysis of the optically-controlled FR, performed
in the framework of the coupled-channel \ model \cite{Nicholson}, suggests
that a laser beam focused on a relatively narrow spot will, generally,
induce a linear component of the potential barrier, in addition to the
nonlinear one introduced above (as demonstrated in recent experimental work
\cite{Marchant}, an effective linear potential produced by a tightly focused
laser beam may produce effects similar to those induced by multi-peak
potentials). However, the linear component is weaker as it is not a resonant
one, and, as shown below, the nonlinear potential barrier produces a much
stronger effect on the operation of the soliton interferometer than its
linear counterpart (irrespective of the particular width of the barrier).
For these reasons, we focus below on the nonlinear-barrier model (\ref{GPE}%
), with $\varepsilon _{0}=0$.

The rest of the paper is structured as follows. First, in Section II we
consider the underlying problem of the scattering of incident waves on the
nonlinear splitter. For the linear plane wave ($g=\Omega =0$ in Eq. (\ref%
{GPE})) we obtain an exact solution, and an approximate analytical one is
obtained for the splitting of an incident soliton ($g=1$). The analysis of
the full model of the interferometer is reported in Section III. The central
issue of the operation of the ``loaded" interferometer (the one with the
target placed in one of its arms), using the nonlinear splitter, is preceded
by the consideration of simpler situations, including revisiting the
interferometer model with the linear splitter, where additional results are
obtained, which help one to compare efficiencies provided by the linear and
nonlinear splitters. These considerations are carried out by means of
systematic simulations, in combination with some analytical approximations.
The most essential result of the work is reported in the last subsection of
Section III: the use of the nonlinear splitter provides for sensitivity of
the soliton-based interferometer which is \emph{definitely superior} to what
is offered by the use of the linear potential barrier. The paper is
concluded by Section IV.

\section{Analytical considerations: scattering on the nonlinear potential
barrier}

\subsection{The exact solution for linear plane waves}

The solution of the scattering problem for the linearized GPE in free space
with the linear $\delta $-shaped potential barrier (Eq. (\ref{GPE}) with $%
g=\Omega =\varepsilon _{2}=0$) is commonly known \cite{Griffiths}:%
\begin{equation}
\phi \left( x,t\right) =Ae^{-i\left( k^{2}/2\right) t}\left\{
\begin{array}{c}
e^{ikx}+re^{-ikx},~\mathrm{at}~~x<0; \\
\tau e^{ikx},~\mathrm{at}~~x>0.%
\end{array}%
\right.  \label{scatt}
\end{equation}%
where $k>0$ and real $A$ are arbitrary wavenumber and amplitude of the
incident wave, the transmission and reflection amplitudes being%
\begin{equation}
\tau (\varepsilon _{0})=-\frac{ik}{\varepsilon _{0}-ik}\equiv i|\tau
|e^{i\theta },~r\left( \varepsilon _{0}\right) =-\frac{\varepsilon _{0}}{%
\varepsilon _{0}-ik}\equiv \left\vert r\right\vert e^{i\theta }  \label{TR}
\end{equation}%
(the representation of the amplitudes in the form of the absolute values and
phases aims to stress the phase shift of $\pi /2$ between them). The result (%
\ref{TR}) can also be applied to the scattering of broad but finite wave
packets, with size%
\begin{equation}
L\gg \pi /k  \label{L}
\end{equation}%
and group velocity $v=k$.

An exact solution can be constructed too for the linearized GPE in free
space with the $\delta $-shaped combined linear-nonlinear potential barrier,
corresponding to Eq. (\ref{GPE}) with $g=\Omega =0$. It is easy to see that,
in this case, Eq. (\ref{TR}) may still be used, with $\varepsilon _{0}$
replaced by
\begin{equation}
\varepsilon _{\mathrm{eff}}\equiv \varepsilon _{0}+\varepsilon
_{2}\left\vert \psi (x=0)\right\vert ^{2}=\varepsilon _{0}+\varepsilon
_{2}A^{2}\left\vert \tau \left( \varepsilon _{\mathrm{eff}}\right)
\right\vert ^{2}=\frac{\varepsilon _{0}\varepsilon _{\mathrm{eff}%
}^{2}+\left( \varepsilon _{0}+\varepsilon _{2}A^{2}\right) k^{2}}{%
\varepsilon _{\mathrm{eff}}^{2}+k^{2}}.  \label{eff}
\end{equation}%
A simple consideration demonstrates that this cubic equation for $%
\varepsilon _{\mathrm{eff}}$ always has a single physically relevant
solution.

In the case of $\varepsilon _{0}=0$ (the purely nonlinear defect, which is
the case of major interest), Eq. (\ref{eff}) simplifies, remaining a cubic
equation:%
\begin{equation}
\varepsilon _{\mathrm{eff}}^{3}+k^{2}\varepsilon _{\mathrm{eff}}=\varepsilon
_{2}A^{2}k^{2}.  \label{simple}
\end{equation}%
The relevant solution of Eq. (\ref{simple}) is a monotonically growing
function of $k^{2}$. In particular, in the limit of $k^{2}\rightarrow 0$,
the solution is%
\begin{equation}
\varepsilon _{\mathrm{eff}}\approx \left( \varepsilon _{2}A^{2}\right)
^{1/3}k^{2/3},  \label{2/3}
\end{equation}%
which means that in this limit the transmission coefficient is%
\begin{equation}
\left\vert \tau \left( \varepsilon _{\mathrm{eff}}\right) \right\vert
^{2}\approx \left( \varepsilon _{2}A^{2}\right) ^{-2/3}k^{2/3},
\end{equation}%
to be compared with a much smaller asymptotic expression, $\left\vert \tau
\left( \varepsilon _{0}\right) \right\vert ^{2}\approx k^{2}/\varepsilon
_{0}^{2}$, in the linear model, see Eq. (\ref{TR}). On the other hand, in
the limit of $k^{2}\rightarrow \infty $, the solution to Eq. (\ref{simple})
is asymptotically constant:%
\begin{equation}
\varepsilon _{\mathrm{eff}}\approx \varepsilon _{2}A^{2},  \label{constant}
\end{equation}%
which makes the situation similar to that in the linear model. For amplitude
$A=1$, the so obtained result, in the form of%
\begin{equation}
T\equiv \left\vert \tau \left( k\right) \right\vert ^{2}=k^{2}/\left(
\varepsilon _{\mathrm{eff}}^{2}+k^{2}\right) ,  \label{Teff}
\end{equation}%
as obtained from the solution of cubic equation (\ref{simple}), is displayed
by the thin continuous line in Fig. \ref{f2}(b).

In the case of $\varepsilon _{0}=0$, $\varepsilon _{2}<0$ (the \emph{%
attractive} nonlinear defect), the respective scattering problem was solved
in Ref. \cite{Azbel}. In that case, it gives rise to a localized
modulational instability of the incident wave, when its amplitude exceeds a
critical value.

\subsection{The scattering of solitons on the nonlinear barrier}

The interaction of the incident soliton with the local defect may be
analyzed by means of the perturbation theory. For the linear potential
barrier, this is a known procedure \cite{old,narrow-barrier}. In the case of
the nonlinear barrier, the effective potential of the interaction of soliton
(\ref{sol})\ with the nonlinear defect is given by the respective term in
the Hamiltonian corresponding to Eq. (\ref{GPE}), that should be added to
energy (\ref{Esol}) of the free soliton:%
\begin{equation}
U_{\mathrm{int}}(\xi )=\left( \varepsilon _{2}/2\right) \left\vert \phi
\left( x=0\right) \right\vert ^{4}=\varepsilon _{2}A^{4}\mathrm{sech}%
^{4}\left( A~\xi \right) ,  \label{Uint}
\end{equation}%
the height of the corresponding potential barrier being
\begin{equation}
U_{0}=\varepsilon _{2}A^{4}/2.  \label{U0}
\end{equation}%
For comparison, the height of the barrier created by the linear $\delta $%
-shaped term in Eq. (\ref{GPE}) is $U_{0}^{(0)}=\varepsilon _{0}A^{2}.$

The kinetic energy of the moving soliton being $E_{\mathrm{kin}}=Av^{2}$,
see Eq. (\ref{Esol}), the critical value of the velocity separating the
rebound of the soliton from the nonlinear defect and its passage is
determined by relation $E_{\mathrm{kin}}=U_{0}$, i.e.,
\begin{equation}
V_{\mathrm{cr}}^{2}=\left( \varepsilon _{2}/2\right) A^{3}.  \label{c-cr}
\end{equation}%
Further if, at $v=v_{\mathrm{cr}}$, the incident soliton, which comes to a
halt around $x=0$, splits into a pair of secondary ones with amplitudes $A/2$
(to provide for the conservation of the total norm, per Eq. (\ref{Esol}))
and velocities $\pm v_{\mathrm{spl}}$, which may be predicted from the
conservation of the total energy. Indeed, it follows from Eq. (\ref{Esol})
that the corresponding energy-balance equation is $-(1/3)A^{3}+Av_{\mathrm{cr%
}}^{2}=2\left[ -(1/3)\left( A/2\right) ^{3}+(A/2)v_{\mathrm{spl}}^{2}\right]
,$ i.e.,%
\begin{equation}
v_{\mathrm{spl}}^{2}=v_{\mathrm{cr}}^{2}-\left( A/2\right) ^{2}=\left(
A^{2}/2\right) \left( \varepsilon _{2}A-1/2\right) .  \label{c-spl}
\end{equation}%
Thus, the efficient splitting is possible if condition $v_{\mathrm{spl}%
}^{2}>0$ holds, as given by Eq. (\ref{c-spl}), i.e., at
\begin{equation}
\varepsilon _{2}>\left( \varepsilon _{2}\right) _{\min }\equiv 1/\left(
2A\right) .  \label{min}
\end{equation}%
It is relevant to compare this result with its counterpart in the case of
the linear potential barrier, which can be easily derived in a similar way: $%
\varepsilon _{0}>\left( \varepsilon _{0}\right) _{\min }\equiv A/4$. As
follows from here, $\left( \varepsilon _{2}\right) _{\min }$ decreases with
the increase of the soliton's amplitude, $A$, while $\left( \varepsilon
_{0}\right) _{\min }$ grows with $A$ (this result for the linear splitter is
similar to one recently reported in Ref. \cite{Helm}).

An alternative interpretation of Eq. (\ref{min}) is possible too: for given
strength $\varepsilon _{2}$ of the nonlinear defect, the incident soliton
will split if its amplitude exceeds a minimum (critical) value,
\begin{equation}
A>A_{\mathrm{c}}=\left( 2\varepsilon _{2}\right) ^{-1}.  \label{0.5}
\end{equation}%
On the contrary, the linear defect with strength $\varepsilon _{0}$ will
split the soliton if its amplitude is not too large:
\begin{equation}
A<A_{\mathrm{c}}^{(0)}=4\varepsilon _{0}.  \label{4}
\end{equation}%
These approximate analytical results are compared with numerical findings
below, see Fig. \ref{f6}.

Furthermore, for the combined linear-nonlinear barrier, Eq. (\ref{c-spl}) is
replaced by%
\begin{equation}
v_{\mathrm{spl}}^{2}=v_{\mathrm{cr}}^{2}-\left( A/2\right) ^{2}=\left(
A^{2}/2\right) \left[ \varepsilon _{2}A+2\left( \varepsilon _{0}/A\right)
-1/2\right] ,
\end{equation}%
hence condition (\ref{min}) is replaced by%
\begin{equation}
2\varepsilon _{2}A^{2}-A+4\varepsilon _{0}>0.  \label{quadr}
\end{equation}%
A simple corollary of Eq. (\ref{quadr}) is that, at $\varepsilon
_{2}\varepsilon _{0}>1/32$, the splitting should be possible for any value
of $A$.

\section{Numerical results}

\subsection{Splitting of the incident Gaussian pulse on the linear and
nonlinear barriers in the linear equation}

First, following the previous section, we briefly consider the scattering of
pulses in the framework of linearized GPE (\ref{GPE}), with $g=0$. In this
case, the ground state of the HO is commonly known, in the absence of the
splitting barrier:
\begin{equation}
\phi =A\exp \left[ \left( \Omega /2\right) \left( it-x^{2}\right) \right] ,
\label{Gauss}
\end{equation}%
where $A$ is an arbitrary amplitude. If placed off the center, this Gaussian
pulse oscillates with frequency $\Omega $.

In simulations, the center of Gaussian (\ref{Gauss}) was initially set at $%
x=-x_{0}<0$ with zero velocity. Accordingly, rolling down in the HO
potential, the pulse impinges upon the splitter with velocity%
\begin{equation}
k=\sqrt{\Omega }x_{0},  \label{k}
\end{equation}%
which is denoted like the wavenumber in the scattering problem, $k$, because
(as mentioned above), for broad pulses satisfying condition (\ref{L}) the
velocity actually plays the role of $k$. In this approximation, the
intensity of the transmitted wave can be obtained from Eq. (\ref{TR}),
\begin{equation}
T_{1}\equiv |\tau |^{2}=k^{2}/(k^{2}+\varepsilon _{0}^{2}),  \label{T1lin}
\end{equation}%
where subscript $1$ implies the first collision. In the numerical scheme,
the $\delta $-function was typically replaced by the rectangular potential
tower of width $\Delta x=0.4$ (then, the above-mentioned estimate $\Delta
X\sim 1$ $\mathrm{\mu }$m for the width measured in physical units implies
the choice of length scale $X_{0}\sim 2.5$ $\mathrm{\mu }$m, see Eq. (\ref%
{scaled})). Most important results were reproduced for other values of $%
\Delta x$ too, to check that they do not essentially depend on $\Delta x$,
see Fig. \ref{f11} below.

\begin{figure}[tbp]
\begin{center}
\includegraphics[height=4.cm]{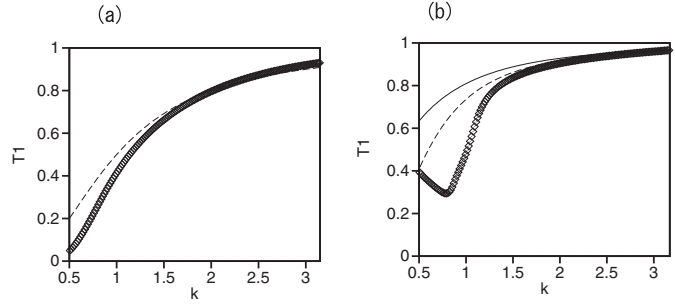}
\end{center}
\caption{The transmission coefficient, $T_{1}$, vs. the collision velocity, $%
k$, for the Gaussian pulse impinging on the linear (a) or nonlinear (b)
potential barrier, in the case of the linearized GPE with $\Omega ^{2}=0.004$%
. (a) $\protect\varepsilon _{1}=1,\protect\varepsilon _{2}=0$; (b) $\protect%
\varepsilon _{1}=0,\protect\varepsilon _{2}=0.6$, and the pulse's amplitude
is $A=1$, see Eq. (\protect\ref{Gauss}). Chains of symbols show numerical
results, while the dashed lines depict the plane-wave approximations (%
\protect\ref{T1lin}) and (\protect\ref{T1nonlin}). The thin continuous line
in (b) separately shows the exact solution for the scattering of the plane
incident wave on the $\protect\delta $-shaped nonlinear splitter, as given
by Eq. (\protect\ref{Teff}).}
\label{f2}
\end{figure}

The simulations demonstrate that the incident Gaussian pulse splits into two
secondary pulses and additional small-amplitude radiation waves. Figure \ref%
{f2}(a) shows $T_{1}$ vs. $k$, as found from the simulations, and compares
it to analytical approximation (\ref{T1lin}). Naturally, the approximation
is accurate for large $k$, and inaccurate for smaller $k,$ which does not
satisfy condition (\ref{L}) (as follows from Eq. (\ref{Gauss}), in the
present notation $\Omega ^{-1/2}$ plays the role of $L$, hence, for $\Omega
=0.004$, adopted in Fig. \ref{f2}, the condition amounts to $k\gg 0.2$).

For the nonlinear splitter, a coarse approximation for the transmission
coefficient is given by Eq. (\ref{T1lin}) with $\varepsilon _{0}$ replaced
by $\varepsilon _{2}A^{2}$, cf. Eq. (\ref{eff}):
\begin{equation}
T_{1}=k^{2}/(k^{2}+\varepsilon _{2}^{2}A^{4})  \label{T1nonlin}
\end{equation}%
Figure \ref{f2}(b) shows that the discrepancy of this approximation at small
$k$ is essentially larger than in the case of the linear splitter. The plots
are not extended to $k<0.5$, as in that case the simulations do not show
clear separation between the slowly moving broad transmitted and reflected
pulses.

\subsection{The operation of the ``idle" interferometer in the linear regime}

Proceeding to modeling the work of the interferometer, with the original
splitting and subsequent recombination, it is natural to start with the
linear model, based on Eq. (\ref{GPE}) with $g=\varepsilon _{2}=0$, $%
\varepsilon _{0}>0$. Note that the target which should be detected by the
interferometer is not introduced yet, therefore we name the setting ``idle".
As above, a shifted Gaussian pulse (\ref{Gauss}) is used as the input.

The outcome of the linear operational cycle may be predicted as the product
of the two scattering events, approximated by amplitudes (\ref{TR}). In the
framework of the linear model, each secondary pulse acquires the same
additional phase in the course of the half-oscillation in the HO trap
between the two events, therefore this phase shift cancels in the analysis.
Thus, the effective transmission and reflection amplitudes for routing the
incident pulse to the right and left are, respectively, $2r\tau =2i|\tau
||r|e^{2i\theta }$ and $\tau ^{2}-r^{2}=(|\tau |^{2}-|r|^{2})e^{2i\theta }$.
Accordingly, the overall transmission coefficient is%
\begin{equation}
T_{2}=|2r\tau |^{2}=4T_{1}(1-T_{1}),  \label{T1T2}
\end{equation}%
where $T_{1}$ is the transmission coefficient for the first collision given
by Eq. (\ref{T1lin}), and relation $|r|^{2}+|\tau |^{2}=1$ is taken into
regard.

\begin{figure}[tbp]
\begin{center}
\includegraphics[height=4.cm]{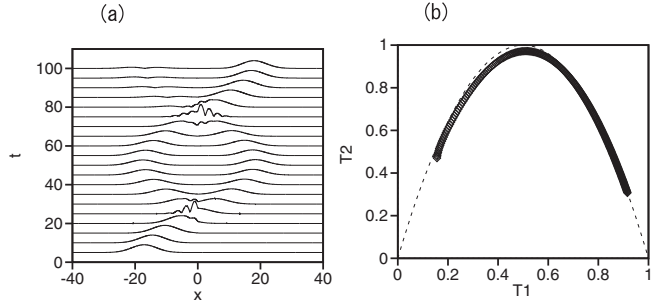}
\end{center}
\caption{(a) The evolution of the Gaussian pulse with the center initially
placed at $x=-18$, in the case of the linearized GPE (\protect\ref{GPE})
with $g=0$, $\Omega ^{2}=0.004$ and $\protect\varepsilon _{0}=1$, $\protect%
\varepsilon _{2}=0$. In this case, Eqs. (\protect\ref{k}) and (\protect\ref%
{T1lin}) yield $T_{1}\approx 0.56$. (b) The relation between the
transmission coefficients corresponding to the primary splitting and
subsequent \ recombination, $T_{1}$ and $T_{2}$, in the same system. The
chain of symbols and dashed line show, respectively, numerical results and
analytical approximation (\protect\ref{T1T2}).}
\label{f3}
\end{figure}

Equation (\ref{T1T2}) predicts $T_{2}=1$ in the case of $T_{1}=1/2$, i.e.,
complete recombination of the secondary pulses, after the second collision,
into the Gaussian pulse moving in the original direction (to the right). The
full simulation of the linear model, displayed in Fig. \ref{f3}(a) for
parameters corresponding to $T_{1}\approx 0.56$ indeed demonstrates
virtually complete merger of the split pulses. In addition, Fig. \ref{f3}(b)
demonstrates that Eq. (\ref{T1T2}) very accurately approximates data
collected from the direct simulations.

\subsection{The operation of the idle soliton-based interferometer with the
linear splitter}

\begin{figure}[tbp]
\begin{center}
\includegraphics[height=4.cm]{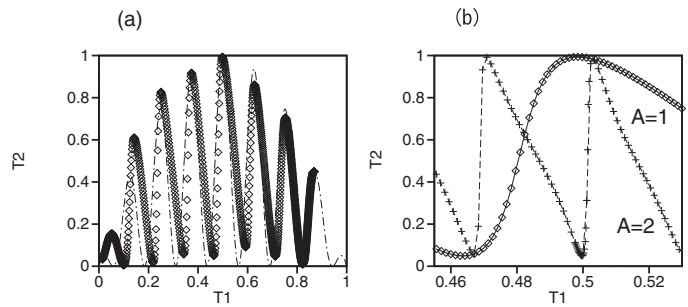}
\end{center}
\caption{(a) The relation between the transmission coefficients
corresponding to the original splitting and subsequent \ recombination of
the soliton, $T_{1}$ and $T_{2}$, in the model of the soliton-based
interferometer (``idle", i.e., without the target to be detected) in the
model based on Eq. (\protect\ref{GPE}) with $\Omega ^{2}=0.004$, $\protect%
\varepsilon _{0}=1$, and $\protect\varepsilon _{2}=0$. The soliton's
amplitude is $A=1,$ see Eq. (\protect\ref{sol}).The chain of symbols and
dashed line show, respectively, numerical results and the analytical
prediction provided by Eq. (\protect\ref{T2T1}). (b) A zoom of (b) around $%
T_{1}=0$, for $A=1$ and $A=2$.}
\label{f4}
\end{figure}

Addressing the splitting and subsequent recombination of soliton (\ref{sol})
in the model based on Eq. (\ref{GPE}) with $g=1$, we expect that a phase
difference, $\Delta \theta =(T_{1}A^{2}-R_{1}A^{2})(\pi /2\Omega )$ is added
to the pair of secondary (split) solitons at the moment of their collision ($%
t=\pi /\Omega $), due to the growth of their phases in time, as $%
(T_{1}A^{2}/2)t$ and $(R_{1}A^{2}/2)t$, after the splitting induced by the
first collision (here, $R_{1}\equiv |r_{1}|^{2}=1-T_{1}$ is the respective
reflection coefficient). Then, the eventual transmission coefficient may be
evaluated as
\begin{equation}
T_{2}=T_{1}(1-T_{1})|1+e^{i\Delta \theta }|^{2}=2T_{1}(1-T_{1})[1+\cos
\{(2T_{1}-1)A^{2}(\pi /2\Omega )\}],  \label{T2T1}
\end{equation}%
which oscillates sinusoidally as a function of $T_{1}$. Figure \ref{f4}(a)
shows the relation between $T_{1}$ and $T_{2}$, as obtained from numerical
simulations. It is compared to the analytical approximation (\ref{T2T1}),
which shows good agreement. Figure \ref{f4}(b) zooms the relation around $%
T_{1}=0.5$. Note that smooth moderately asymmetric oscillations observed for
$A=1$ are replaced by strongly asymmetric sawtooth-like oscillations for $%
A=2 $, which corresponds to stronger background nonlinearity. Figure \ref{f3}
clearly identifies a point of the virtually perfect recombination ($T_{2}=1$%
) very close to $T_{1}=0.5$, and a series of satellite points with gradually
deteriorating recombination quality ($T_{2}<1$).

\begin{figure}[tbp]
\begin{center}
\includegraphics[height=4.cm]{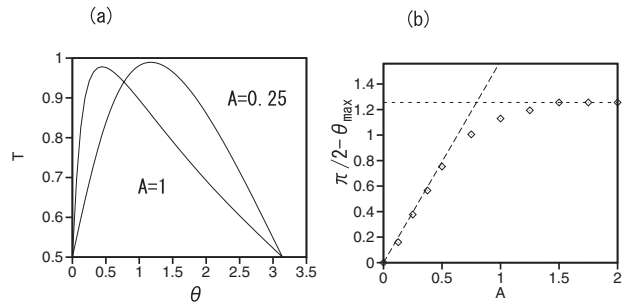}
\end{center}
\caption{(a) The relation between the recombination rate, $T$, and the phase
shift, $\protect\theta $, of the two colliding solitons, taken per Eq. (%
\protect\ref{sechsech}) with amplitudes $A=0.25$ and $1$, as obtained from
simulations of Eq. (\protect\ref{GPE}) with $g=1$, $\Omega ^{2}=0.004$, $%
\protect\varepsilon _{0}=1,$ $\protect\varepsilon _{2}=0$. (b) $\protect\pi %
/2-\protect\theta _{\max }$ vs. $A$, where $\protect\theta _{\max }$ is the
phase shift in input (\protect\ref{sechsech}) which produces the largest
recombination rate. The dashed straight line in (b) is a linear fit for
small $A$, $\protect\pi /2-\protect\theta _{\max }\approx 1.58A$.}
\label{f5}
\end{figure}

To understand the asymmetric behavior observed in Fig. \ref{f4}(b) in the
case of the strong nonlinearity, we have performed numerical simulations of
Eq. (\ref{GPE}) for the collision of two solitons with a phase shift, $%
\theta $, taking the initial condition as
\begin{equation}
\phi =A\left\{ \mathrm{sech}\left[ A(x+x_{0})\right] +e^{i\theta }\mathrm{%
sech}\left[ A(x-x_{0})\right] \right\} .  \label{sechsech}
\end{equation}%
Figure \ref{f5}(a) shows the corresponding recombination rate (defined as
the integral transmission coefficient toward $x>0$, $T=\int_{0}^{\infty
}|\phi (x)|^{2}dx/\int_{-\infty }^{+\infty }|\phi (x)|^{2}dx$), calculated
at some moment of time after the end of the complete or incomplete
collision-induced recombination. It is seen that, if $A$ is sufficiently
small, a smooth quasi-sinusoidal dependence of $T$ on $\theta $ is observed,
which changes to sawtooth-like oscillations as $A$ increases, i.e., the
nonlinearity gets stronger. The quasi-sinusoidal behavior is due to the
linear combination of the reflected and transmitted waves in the
nearly-linear regime,
\begin{equation}
T\sim \left (1/2)\{\vert i|\tau |+|r|e^{i\theta }\right\vert
^{2}\}=(1/2)+|r||\tau |\sin \theta ,  \label{sin}
\end{equation}%
see Eq. (\ref{TR}). Phase $\theta _{\max }$ corresponding to the largest
recombination rate, which is $\theta =\pi /2$ in Eq. (\ref{sin}), deviates
from $\pi /2$ as $A$ increases, due to the nonlinearity-induced phase shift.
Figure \ref{f5}(b) shows the deviation, $\pi /2-\theta _{\max }$, as a
function of $A$.

The steep dependence of the recombination rate on $\theta $ is promising for
the operation of the interferometer, with the target placed into one of its
arms, as such a dependence may be used to secure high sensitivity of the
operation, see below. It is relevant to mention that a similar transition
from a smooth dependence to steep one with the increase of the nonlinearity
strength was reported, for a soliton interferometer with a linear splitter,
in Ref. \cite{Martin} (see Figs. 1(d,e) in that work). Thus, Fig. \ref{f5}%
(a) suggests that the accuracy of the interferometer using the linear
splitter should improve if heavier solitons are used, with larger $A$. On
the other hand, the increase of $A$ is limited by the fact that the solitons
with the amplitude exceeding the critical value $A_{\mathrm{c}}^{(0)}$,
given by Eq. (\ref{4}) (or, strictly speaking, by its numerically generated
counterpart), may not split at all. In particular, for $\varepsilon _{0}=1$
considered above, Eq. (\ref{4}) suggests that only values $A<4$ may be
usable.

\subsection{The operation of the idle soliton-based interferometer with the
nonlinear splitter}

\begin{figure}[tbp]
\begin{center}
\includegraphics[height=4.cm]{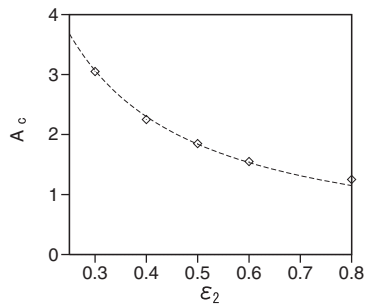}
\end{center}
\caption{The critical amplitude of the incident soliton, $A_{\mathrm{c}}$,
above which it is split by the collision with the nonllinear potential
barrier of strength $\protect\varepsilon _{2}$, see Eq. (\protect\ref{GPE}).
The dashed line shows the fit given by Eq. (\protect\ref{0.92}).}
\label{f6}
\end{figure}

Prior to the simulations of the full soliton-based interferometer model
using the nonlinear splitter, with $\varepsilon _{0}=0$ and $\varepsilon
_{2}>0$ in Eq. (\ref{GPE}), it makes sense to study, in some detail, the
primary collision of soliton (\ref{sol}) with the nonlinear potential
barrier. An essential prediction of the analysis reported in the previous
section is that collision will lead to splitting of the incident soliton if
its amplitude exceeds the critical value, which is given by Eq. (\ref{0.5})
(on the contrary to the case of the linear barrier, which splits the soliton
if its amplitude is smaller than the corresponding critical value, see Eq. (%
\ref{4})). Simulations corroborate the prediction, and produce the critical
value, $A_{\mathrm{c}}$, which is displayed as a function of $\varepsilon
_{2}$ in Fig. \ref{f6}. The numerically found dependence may be fitted to
\begin{equation}
A_{\mathrm{c}}=0.92/\varepsilon _{2},  \label{0.92}
\end{equation}%
which yields larger values than the analytical estimate (\ref{0.5}), but the
dependence on $\varepsilon _{2}$ is essentially the same as predicted. The
discrepancy in the overall factor ($0.92$ versus $0.5$) is explained by the
fact that analytical consideration did not take into account deformation of
the soliton's shape in the course of the splitting.

\begin{figure}[t]
\begin{center}
\includegraphics[height=4.cm]{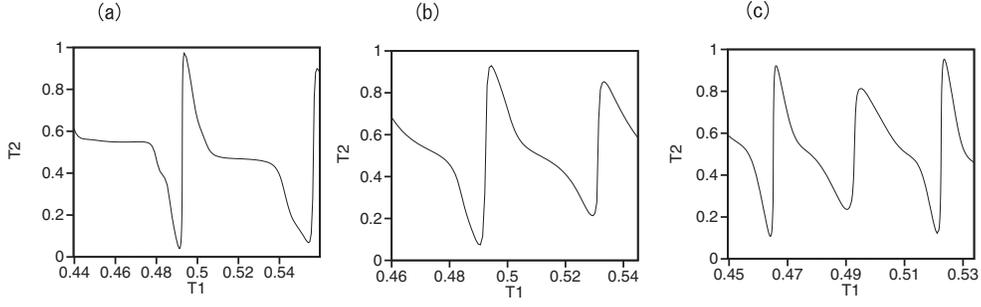}
\end{center}
\caption{Numerically generated relations between the transmission
coefficients corresponding to the original splitting and subsequent \
recombination of the soliton, $T_{1}$ and $T_{2}$ (cf. Fig. \protect\ref{f3}%
), in the model of the soliton-based interferometer (idle, i.e., without the
target to be detected), in the model based on Eq. (\protect\ref{GPE}) with $%
\Omega ^{2}=0.004$ and the nonlinear splitter: $\protect\varepsilon _{0}=0$,
$\protect\varepsilon _{2}=0.6$. The amplitude of the incident soliton is $%
A=1.7$ in (a), $A=2$ in (b), and $A=2.3$ in (c).}
\label{f7}
\end{figure}

Proceeding to modeling the full scheme (but still for the interferometer in
the idle mode), Fig. \ref{f7} shows relations between $T_{2}$ and $T_{1}$,
similar to those displayed in Fig. \ref{f4} for the interferometer with the
linear splitter, at different values of the amplitudes. The smallest one is
chosen as $A=1.7$ because Eq. (\ref{0.92}) shows that the incident soliton
may be split by the nonlinear barrier with $\varepsilon _{2}=0.6$,
considered here, only for $A>\allowbreak 1.\,\allowbreak 53$. Sharp sawtooth
oscillations are observed in all the cases. The period of the oscillations
increases with $A$, similar to what was seen for the model with the linear
splitter. However, on the contrary to that model, in which the asymmetry and
sharpness monotonically increase with $A$, here they are largest for the
smallest amplitude considered, $A=1.7$. This finding is naturally explained
by the fact that the variation must indeed be steepest closer to critical
point. Because the steepest variation suggests the highest sensitivity, the
use of the nonlinear splitter is potentially more promising than of its
linear counterpart. Also promising is the fact that a relatively light
soliton, which is easier to make in the experiment, will provide the higher
accuracy. On the other hand, quantum fluctuations may come into the play for
very light solitons \cite{quantum,Helm}.

\begin{figure}[tbp]
\begin{center}
\includegraphics[height=4.cm]{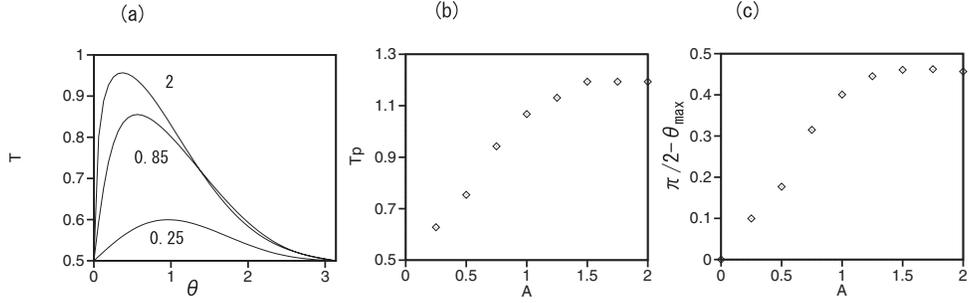}
\end{center}
\caption{(a) The relation between the recombination rate, $T$, and the phase
shift, $\protect\theta $, of the two colliding solitons, taken per Eq. (%
\protect\ref{sechsech}) with amplitudes $A=0.25$, $0.85$, and $2$, as
obtained from simulations of Eq. (\protect\ref{GPE}) with $g=1$, $\Omega
^{2}=0.004$, $\protect\varepsilon _{0}=0,$ $\protect\varepsilon _{2}=0.6$.
Value $A=0.85$ corresponds to Fig.~7(a). (b) The largest (peak) value, $T_{%
\mathrm{p}}$, of the recombination rate, which may be attained at given $A$,
vs. $A$. (c) $\protect\pi /2-\protect\theta _{\max }$ vs. $A$, where $%
\protect\theta _{\max }$ is the phase shift in input (\protect\ref{sechsech}%
) which produces the largest recombination rate.}
\label{f8}
\end{figure}

As suggested by the above analysis of the scheme with the linear potential
barrier (see Fig. \ref{f5}), the operation of the setup with the nonlinear
splitter should be further characterized by the consideration of the
on-splitter collision of two solitons with the same amplitude $A$ and phase
shift $\theta $, corresponding to input (\ref{sechsech}). Figure \ref{f8}
displays numerically generated dependences of the recombination rate, $T$,
on $\theta $, for several values of the amplitude. It is seen that the
largest value of $T$ increases with $A$, as well the deviation of the phase
shift, $\theta _{\max }$, providing the largest value, from $\pi /2$.\ These
dependences are markedly different from their counterparts in the model with
the linear splitter, cf. Fig. \ref{f5}.

\subsection{The operation of the loaded soliton-based interferometer}

The most important step of the analysis is its application to the model of
the interferometer in which one arm is ``loaded" with the target that should
be detected by the device. The above results, which demonstrate strong
dependence of the recombination on phase changes suggest that the detection
procedure may be quite sensitive.

The model of the loaded interferometer is derived from Eq. (\ref{GPE}) (with
$g=1$) by adding the target in the form of the $\delta $-functional linear
potential placed at $x=x_{0}/2$, with strength $\varepsilon _{3}$:
\begin{equation}
i\frac{\partial \phi }{\partial t}=\left[ -\frac{1}{2}\frac{\partial ^{2}}{%
\partial x^{2}}-|\phi |^{2}+\frac{1}{2}\Omega ^{2}x^{2}+\varepsilon
_{0}\delta (x)+\varepsilon _{2}\delta (x)|\phi |^{2}+\varepsilon _{3}\delta
\left( x-\frac{x_{0}}{2}\right) \right] \phi .  \label{loaded}
\end{equation}%
Note that $\varepsilon _{3}$ may be both positive and negative (repulsive or
attractive), unlike $\varepsilon _{0}$ and $\varepsilon _{2}$ which should
be positive to work as splitters (in principle, incident solitons hitting a
local potential well may feature splitting too \cite{Brand}). Simulations of
this model aimed to produce the recombination rate as a function of $%
\varepsilon _{3}$ for different values of the soliton's amplitude, $A$.
Here, we present results obtained for the case when the largest (peak) value
of the recombination rate (as above, it is defined as the transmission
coefficient, $T_{2}$, produced by the collision of the secondary solitons)
is close to $T_{\mathrm{p}}=0.5$ at $\varepsilon _{3}=0$, which actually
implies the choice of parameters at which the recombination does not occur
in the idle interferometer.

\begin{figure}[t]
\begin{center}
\includegraphics[height=4.cm]{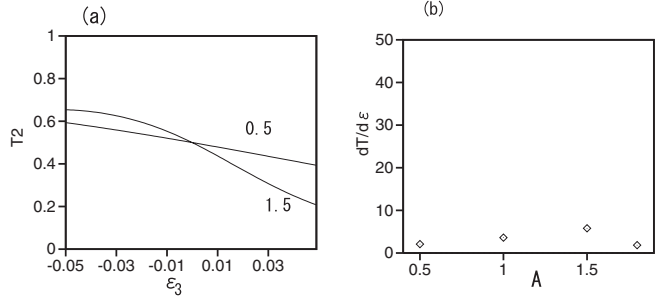}
\end{center}
\caption{(a) The dependence of the recombination rate, $T_{2}$, on the
target's strength, $\protect\varepsilon _{3},$ for two values of the initial
amplitude of the probe soliton, $A=$ $0.5$ and $1.5$, produced by
simulations of Eq. (\protect\ref{loaded}) with the\emph{\ linear} splitter, $%
\protect\varepsilon _{0}=1,\protect\varepsilon _{2}=0$. (b) The dependence
of the sensitivity, $dT_{2}/d\protect\varepsilon _{3}$ at $\protect%
\varepsilon _{3}=0$, for the same system. }
\label{f9}
\end{figure}

First, Fig. \ref{f9}(a) shows the dependence of the recombination rate on
the target's strength, $\varepsilon _{3}$, in the interferometer using the
linear splitter with $\varepsilon _{0}=1,\varepsilon _{2}=0$. It is observed
that $T_{2}$ varies very smoothly at $A=0.5$. To quantify the sensitivity,
we have defined it as the value of $dT_{2}/d\varepsilon _{3}$ at $%
\varepsilon _{3}=0$. Figure \ref{f9}(b) shows its dependence on the initial
amplitude of the probe soliton.
\begin{figure}[tbp]
\begin{center}
\includegraphics[height=4.5cm]{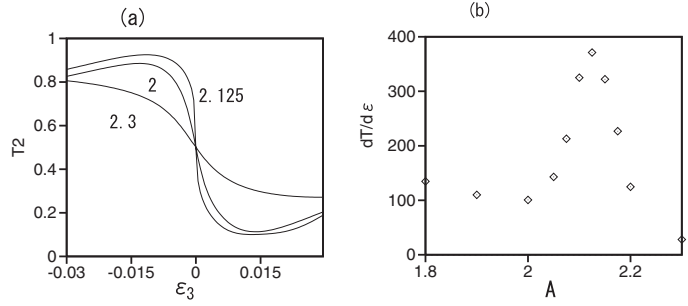}
\end{center}
\caption{The same as in Fig. \protect\ref{f9}, but for the system with the
\emph{nonlinear splitter} ($\protect\varepsilon _{0}=0,\protect\varepsilon %
_{2}=0.6$), at three values of the initial amplitude of the probe soliton: $%
A=2.0,2.125$, and $2.3$. }
\label{f10}
\end{figure}

Next, Fig. \ref{f10}(a) shows the most essential results produced by the
analysis, i.e., $T_{2}$ as a function of $\varepsilon _{3}$, and the
sensitivity, $dT_{2}/d\varepsilon _{3}$ at $\varepsilon _{3}=0$, as a
function of the initial soliton's amplitude, $A$ in the system using the
\emph{nonlinear splitter}, with $\varepsilon _{0}=0,\varepsilon _{2}=0.6$.
In this case, $T_{2}$ varies \emph{much faster} near $\varepsilon _{3}=0$,
and the sensitivity is \emph{much higher} in comparison with the setup using
the linear splitter, cf. Fig. \ref{f9}. In particular, the sensitivity takes
a very large value $370$ near $A=2.125$, as shown in Fig. \ref{f10}(b). A
relative width of the high-sensitivity region is $\Delta A/A\simeq 0.03$,
implying that the optimal use of the present setup requires rather accurate
selection of parameters of the probe soliton, which may be a challenge to
the implementation of the present scheme. In principle, if a stable source
of solitons is available, such as a matter-wave soliton laser \cite{laser},
the scheme may be adjusted to the optimal operation regime empirically, by
tuning parameters of the optical beam which controls the action of the
nonlinear splitter. However, a discussion of further details of the
experimental implementation does not seem relevant in this theoretical paper.

\begin{figure}[tbp]
\begin{center}
\includegraphics[height=4.5cm]{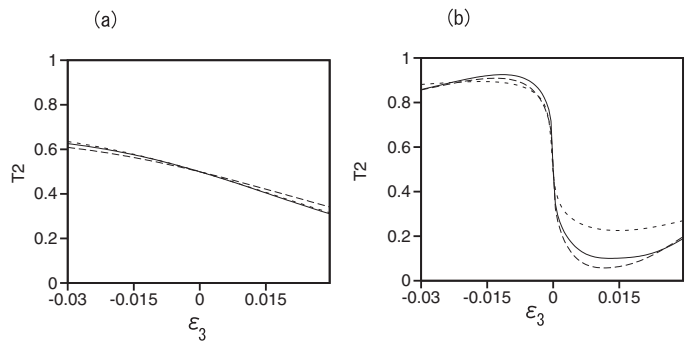}
\end{center}
\caption{(a) and (b): The same as in Figs. \protect\ref{f9}(a) for $A=1.5$
and \protect\ref{f10}(a) for $A=2.125$, respectively, but with widths $%
\Delta x=0.2$ (long-dash lines), $0.4$ (solid lines), and $0.6$ (short-dash
lines) of the finite barrier approximating the $\protect\delta $-function.}
\label{f11}
\end{figure}

While the highest sensitivity is attained at $\varepsilon _{2}=0.6$, similar
results were obtained for other values of strength $\varepsilon _{2}$, with
lower values of the largest sensitivity, which is $26$ at $\varepsilon
_{2}=0.7$, and $102$ at $\varepsilon _{2}=0.5$. It was also checked that the
results reported in this subsection are robust in the sense that they do not
vary conspicuously with the change of width $\Delta x$ used for the
approximation of the $\delta $-function. The robustness is illustrated by
Fig. \ref{f11}, which demonstrates that findings collected in Figs. \ref{f9}%
(a) and \ref{f10}(a) vary weakly with $\Delta x$. Thus, the use of the
nonlinear splitter offers the possibility to build very efficient
soliton-based interferometers, in comparison with the previously developed
setups, that use the linear splitter.

\section{Conclusion}

We have introduced the model of the soliton-based interferometer which
utilizes, unlike the previously studied schemes, the nonlinear splitter, in
the form of the localized region with the strongly repulsive intrinsic
nonlinearity, embedded into the uniform self-attractive medium. It was
demonstrated that this setting may be realized with the help of the Feshbach
resonance, controlled by a laser beam focused on a narrow region, where it
creates the repulsive nonlinearity. The systematic analysis of the
scattering of plane waves and solitons on the localized nonlinear potential,
and of the operation of the full interferometric setup, has been carried out
by means of combined analytical and numerical methods. For the sake of
comparison with the new setup, additional analysis was also developed for
the traditional one, based on the linear splitter. Essential results include
the exact solution for the scattering of the plane wave in the linear medium
on the nonlinear $\delta $-functional nonlinear potential and perturbative
analysis of the splitting of the incident solitons by the same potential.
The most significant finding is that the use of the nonlinear splitter
predicts operational regimes for the interferometer with the sensitivity to
the target much higher than provided, in the same range of parameters, by
the usual linear splitter.

The work may be extended by considering the generalized form of the GPE
which takes into account deviations from the one-dimensionality, making use
of approaches developed in Refs. \cite{1D3D} and \cite{Cuevas}. Another
interesting possibility is the use of probe solitons with embedded
vorticity, cf. Ref. \cite{Luca-vortex}.

\section*{Acknowledgments}

We appreciate valuable discussions with T. C. Killian, M. Olshanii, and R.
G. Hulet. This work was supported, in a part, by the Binational Science Foundation
(US-Israel) through grant No. 2010239. B.A.M. appreciates hospitality of the
Interdisciplinary Graduate School of Engineering Sciences at the Kyushu
University (Fukuoka, Japan).

\end{document}